\documentclass[11pt]{article}%
\usepackage{amsmath}
\usepackage{amssymb}
\usepackage{amsfonts}

\usepackage{cite}
\usepackage{graphicx}
\usepackage{float}
\usepackage{longtable}
\usepackage[utf8]{inputenc}
\usepackage{hyperref}

\usepackage{algorithmicx}
\usepackage[ruled,vlined]{algorithm2e}
\SetAlgoCaptionSeparator{ }

\usepackage{fullpage}

\newcommand{\fref}[1]{Fig.~\ref{#1}}

\newcommand{\sref}[1]{Section~\ref{#1}}

\newenvironment{algo}[1][!h]
  {
   \begin{algorithm}[#1]%
  }{\end{algorithm}}
\providecommand{\U}[1]{\protect\rule{.1in}{.1in}}
\begin{document}

\title{Entanglement Access Control for the Quantum Internet}
%\author[1]{Draft}
%\author[1]{Laszlo Gyongyosi\footnote{L. Gyongyosi is with the School of Electronics and Computer Science, University of Southampton, Southampton SO17 1BJ, U.K., also with the Department of Networked Systems and Services, Budapest University of Technology and Economics, 1117 Budapest, Hungary, and also with the MTA-BME Information Systems Research Group, Hungarian Academy of Sciences, 1051 Budapest, Hungary (e-mail: \href{mailto:l.gyongyosi@soton.ac.uk}{l.gyongyosi@soton.ac.uk})}}

%\author[1]{Laszlo Gyongyosi\footnote{School of Electronics and Computer Science, University of Southampton, Southampton, SO17 1BJ, UK, Email: \href{mailto:l.gyongyosi@soton.ac.uk}{l.gyongyosi@soton.ac.uk}}}
%\affil[1]{School of Electronics and Computer Science, University of Southampton, Southampton, SO17 1BJ, UK}
%\affil[2]{Department of Networked Systems and Services, Budapest University of Technology and Economics, Budapest, H-1117 Hungary}
%\affil[3]{MTA-BME Information Systems Research Group, Hungarian Academy of Sciences, Budapest, H-1051 Hungary}

\author{Laszlo Gyongyosi\thanks{School of Electronics and Computer Science, University of Southampton, Southampton SO17 1BJ, U.K., and Department of Networked Systems and Services, Budapest University of Technology and Economics, 1117 Budapest, Hungary, and MTA-BME Information Systems Research Group, Hungarian Academy of Sciences, 1051 Budapest, Hungary.}
\and Sandor Imre\thanks{Department of Networked Systems and Services, Budapest University of Technology and Economics, 1117 Budapest, Hungary.}}

\date{}

\maketitle
\begin{abstract}
Quantum entanglement is a crucial element of establishing the entangled network structure of the quantum Internet. Here we define a method to achieve controlled entanglement access in the quantum Internet. The proposed model defines different levels of entanglement accessibility for the users of the quantum network. The path cost is determined by an integrated criterion on the entanglement fidelities between the quantum nodes and the probabilities of entangled connections of an entangled path. We reveal the connection between the number of available entangled paths and the accessible fidelity of entanglement and reliability in the end nodes. The scheme provides an efficient model for entanglement access control in the experimental quantum Internet.
\end{abstract}
%\begin{keywords}
%quantum Internet - quantum repeater - quantum entanglement - quantum Shannon theory.
%\end{keywords}

\section{Introduction}
\label{sec1}
In the quantum Internet, the quantum nodes share quantum entanglement among one other, which provides an entangled ground-base network structure for the various quantum networking protocols \cite{ref1,ref2,ref3,ref4,ref5,ref6,ref7,ref8,ref9,ref10,ref11,ref12,ref13,ref14,ref15,ref16,ref17,ref18,ref19,ref20}. In a quantum Internet scenario, the aim of the quantum repeater elements is to extend the range of entanglement through several steps \cite{ref20,ref21,ref22,ref23,ref24,ref25,ref26,ref27,ref28,ref29,ref30,ref31,ref32,ref33}. The available entanglement at the end points has several critical parameters, most importantly the fidelity of the established entanglement (fidelity of entanglement \cite{a3,a4}) and the probability of the existence of a given entangled connection \cite{ref1,ref8}. In an experimental setting, these critical parameters are time-varying since the noise of local quantum memories that store the shared entanglement in the quantum nodes evolves over time, and the probability of entangled connections (shared entanglement between a node pair) also changes dynamically \cite{ref1,ref5,ref6,ref34,ref35,ref36,ref37,ref38,ref39,ref40,ref41,ref42,ref43,a1}. 

In the quantum Internet, several entangled paths (paths formulated by several entangled connections) could exist between a given source-target quantum node pair \cite{ref1,ref5,ref6,ref43,ref44,ref45,ref46,ref47,ref48,ref49,ref50,ref51,ref52,ref53,ref54,ref55,ref56,ref57,ref58,ref59,ref60}. This fact allows us to introduce a method that utilizes this multipath property to change these critical parameters via the number of entangled paths associated with a given end-to-end node pair: the available fidelity of entanglement and the probability of an entangled connection. The model utilizes the reliability (probability) of the entangled connections and the entanglement fidelity coefficient as primary metrics. The decomposition is motivated by the fact that a maximization of the entanglement throughput (number of transmitted Bell states per a given time unit at a particular fidelity) parameter requires also the maximization of the connection probability, and the entanglement fidelity. 

In this work, we define a method for entanglement access control in entangled quantum networks. The entanglement differentiation is achieved via a controlled variability of entanglement fidelity and entangled connection probability between source and target quantum nodes in a quantum repeater network. The proposed approach allows us to define different priority levels of entanglement access for the legal users of the quantum network with respect to the number of available paths. The number of available paths injects an additional degree of freedom to the quantum network, allowing for the selection of the entanglement fidelity and connection reliability for the end nodes. In a straightforward application of our method, the high-priority demands are associated with high fidelity and high connection probability in the end nodes of the user, while the lower-priority users get lower fidelity and lower connection probability in their end nodes. To achieve the differentiation, we define the appropriate cost and path cost functions and the criteria regarding the entanglement fidelity and connection probability for the quantum nodes and entangled connections of the entangled path. The entanglement differentiation utilizes a different number of paths between the source and target nodes allowing a distinction to be made between single-path and multipath scenarios. In a single-path setting, only one entangled path exists between source and target nodes, and therefore the fidelity of entanglement and the probability of existence of the entangled connections between the end nodes are determined only by the nodes of the given entangled path. In a multipath scenario, more than one path exists from a source to a target. 

In our model, a given criterion regarding the entanglement fidelity of the local nodes has to be satisfied for all node pairs on the path referred to as integrated connection probability and fidelity criterion for all entangled connections of the entangled path. The integrated criterion allows us to reach a given entanglement fidelity and a given connection probability between the end nodes of the quantum network.

Our solution utilizes time-varying parameters since the cost functions deal with the evolution of the entanglement fidelity parameter and the connection probabilities, which evolve over time. We define the entanglement access control algorithm for an arbitrary topology quantum repeater network. We also reveal the computational complexity of the method. 

The novel contributions of our manuscript are as follows:

\begin{enumerate}
\item \textit{We define a method to achieve controlled entanglement accessibility in the quantum Internet.} 

\item \textit{The algorithm defines entangled paths between the source and target nodes in function of a particular path cost function.} 

\item \textit{The path cost is determined by an integrated criterion on the entanglement fidelity and the probability of entangled connection.}

\item \textit{The proposed scheme has moderate complexity, providing an efficient entanglement accessibility differentiation, allowing for the construction of different priority levels of entanglement accessibility for users.} 

\item \textit{The results can be straightforwardly applied in the entangled quantum networks of the quantum Internet.}  
\end{enumerate}

This paper is organized as follows. In \sref{sec2}, the basic components of the model are summarized. In \sref{sec3}, the entanglement accessibility methods are discussed. \sref{sec4} proposes the integrated criterion related to entanglement fidelity. \sref{sec5} defines the entanglement access control algorithm. Finally, \sref{sec6} concludes the results.

\section{System Model}
\label{sec2}
\subsection{Entangled Network}
The quantum Internet setting is modeled as follows \cite{ref8}. Let $V$ refer to the nodes of an entangled quantum network $N$, with a transmitter quantum node $A\in V$, a receiver quantum node $B\in V$, and quantum repeater nodes $R_i\in V$, $i=1,\dots ,q$. Let $E=\left\{E_j\right\}$, $j=1,\dots ,m$, refer to a set of edges between the nodes of $V$, where each $E_j$ identifies an $\text{L}_l$-level entangled connection, $l=1,\dots ,r$, between quantum nodes $x_j$ and $y_j$ of edge $E_j$, respectively. The entanglement levels of the entangled connections in the entangled quantum network structure are defined as follows.

\subsubsection{Entanglement Levels in the Quantum Internet}
In a quantum Internet setting, an $N=\left(V,E\right)$ entangled quantum network consists of single-hop and multi-hop entangled connections, such that the single-hop entangled nodes\footnote{The $l$-level entangled nodes ${x,y}$ refer to quantum nodes $x$ and $y$ connected by an entangled connection ${\text{L}_l}$.} are directly connected through an $\text{L}_1$-level entanglement, while the multi-hop entangled nodes communicate through $\text{L}_l$-level entanglement. Focusing on the doubling architecture \cite{ref1,ref5,ref6} in the entanglement distribution procedure, the number of spanned nodes is doubled in each level of entanglement swapping (entanglement swapping is applied in an intermediate node to create a longer distance entanglement \cite{ref1}). Therefore, the $d{\left(x,y\right)}_{\text{L}_l}$ hop distance in $N$ for the $\text{L}_l$-level entangled connection between $x,y\in V$ is denoted by \cite{ref8}
\begin{equation} \label{eq1} 
d{\left(x,y\right)}_{\text{L}_l}=2^{l-1}, 
\end{equation} 
 with $d{\left(x,y\right)}_{\text{L}_l}-1$ intermediate quantum nodes between $x$ and $y$. Therefore, $l=1$ refers to a direct entangled connection between two quantum nodes $x$ and $y$ without intermediate quantum repeaters, while $l>1$ identifies a multilevel entanglement.

\subsubsection{Entanglement Fidelity}
Let
\begin{equation}\label{bell}
{| \beta _{00}  \rangle} ={\textstyle\frac{1}{\sqrt{2} }} \left({\left| 00 \right\rangle} +{\left| 11 \right\rangle} \right)
\end{equation}
be the target Bell state subject to be created at the end of the entanglement distribution procedure between a particular source node $A$ and receiver node $B$. The entanglement fidelity $F$ at an actually created noisy quantum system $\sigma $ between $A$ and $B$ is
\begin{equation}\label{fid}
F\left( \sigma  \right)=\langle  {{\beta }_{00}} |  \sigma |{{\beta }_{00}} \rangle ,
\end{equation}
where $F$ is a value between $0$ and $1$, $F=1$ for a perfect Bell state and $F<1$ for an imperfect state. 
The $F$ entanglement fidelity represents the accuracy of our information about a quantum state \cite{ref1,ref5,ref6}. The fidelity in \eqref{fid} measures the amount of overlap between the ${\left| \beta _{00}  \right\rangle}$ target state \eqref{bell} and the density matrix $\sigma $ that represents our system.
In the entanglement distribution procedure, the usage of the $F$ entanglement fidelity metric rather than other correlation measure functions (concurrence, negativity, quantum discord, quantum coherent information, etc.) \cite{ref4} is motivated by the fact that the fidelity of entanglement is an improvable parameter in a practical setting. The improvement of the fidelity is realizable by the so-called entanglement purification process \cite{ref1}. The entanglement purification takes imperfect entangled states and outputs a higher-fidelity entangled system. Without loss of generality, in an experimental quantum Internet setting, an aim is to reach $F\ge 0.98$ over long distances \cite{ref1,ref5,ref6}.

\subsubsection{Practical Implementation}
An experimental quantum network refers to a set of source users (quantum nodes), destination users (quantum nodes), several intermediate quantum repeaters between them, and to a set of physical node-to-node connections between the physical nodes (the physical attributes of the $l=1$ level connections identify the physical attributes of the physical links between the neighboring nodes). A quantum node is a quantum device with internal quantum memory $\mathcal{M}$, and with the capability of performing local operations (such as the internal processes connected to entanglement purification, entanglement swapping, error correction, etc.)\cite{ref25,ref34,ref35,ref36,ref37,ref38,ref39,ref40,ref41,ref42,ref43,ref44,ref45,ref46,ref47,ref48,ref49,ref50,ref51,ref52,ref53,ref54,ref55,ref56,ref57,ref58,ref59,ref60 }. In a practical setting, the node-to-node entanglement distribution can be implemented by an optical fiber network or via a wireless optical system (free-space channel \cite{a5} or a quantum-based satellite communication channel \cite{ref27}). A physical link $\mathcal{N}$ is characterized by a particular link loss $\mathcal{L}\left( \mathcal{N} \right)$. For a standard-quality optical fiber $\mathcal{N}$, the average link loss is $\mathcal{L}\left( \mathcal{N} \right)\approx 3.4\text{ dB}$, while the maximum of the tolerable link loss for an optical fiber system is $\mathcal{L}\left( \mathcal{N} \right)\approx 4.3\text{ dB}$ \cite{ref1,ref6}. 

In an practical entangled quantum network, the $l>1$ level entangled connections refers to the case when the source and target quantum nodes are not directly connected by a physical link, but by an entangled connection that spans several quantum repeaters. An $l>1$ level entangled connection is formulated by several node-to-node interactions through the physical links in the physical layer.

\section{Entanglement Access}
\label{sec3}
\subsection{Entanglement Fidelity Criterion}

First, we characterize the entanglement fidelity criterion for a given node pair. Using the criterion, we then derive the probability of the existence of single-path and multipath sets with $m$ end-to-end connection-disjoint entangled paths \cite{ref61,ref62} between source and target nodes. The end-to-end connection-disjoint entangled paths share no any common entangled connection between a source node $A$ and a receiver node $B$.

A given entangled connection ${\rm L}_{l} $ is characterized by a particular fidelity $F^{{\rm *}} $, whose quantity classifies the entangled connection, such that $F^{{\rm *}} \ge F_{crit} $, where $F_{crit} $ is a critical lower bound on the fidelity of entanglement. 

Let $E\left(x,y\right)$ refer to the entangled connection between a node pair $\left(x,y\right)$, and let $F_{\Delta } \left(x,y\right)$ be the difference of the fidelity of entanglement in quantum nodes $x$ and $y$, as 
\begin{equation} \label{1)} 
F_{\Delta } \left(x,y\right)=\left|F_{x} -F_{y} \right|<\hat{F}_{\Delta } ,                                          
\end{equation} 
where $\hat{F}_{\Delta } $ is a maximal allowed fidelity distance, $F_{x} \ge F_{crit} $, and $F_{y} \ge F_{crit} $. Since the entangled connections are assumed to be time-varying in the network \cite{ref61,ref62}, the probability that $F_{\Delta } \left(x,y\right)<\hat{F}_{\Delta } $ holds at a given time $t$ for an entangled connection $E_{{\rm L}_{l} } \left(x,y\right)$ is as
\begin{equation} \label{2)} 
\Pr \left(F_{\Delta } \left(x,y\right)<\hat{F}_{\Delta } \right)=\int\limits_{0}^{\hat{F}_{\Delta } }\delta \left(z\right)dz ,                                      
\end{equation} 
where $\delta \left(F_{\Delta } \left(x,y\right)\right)$ is the probability density function of entanglement fidelity distance.

\subsubsection{Single-Path Entanglement Accessibility}

Let ${\rm {\mathcal P}}^{S} $ refer to a single path between $A_{\rho ,U_{k} } $ and $B_{\rho ,U_{k} } $, $k=1,\ldots ,K$, where $\rho $ is a demand, $A_{\rho ,U_{k} } $ and $B_{\rho ,U_{k} } $ are the sender and destination nodes associated with the demand $\rho $ of user $U_k$, $K$ is the number of users. The single entangled path setting means that entanglement can be distributed from $A_{\rho ,U_{k} } $ to $B_{\rho ,U_{k} } $ through only one given path in the network $N$. Let it be assumed that ${\rm {\mathcal P}}^{S} $ consists of $g$ entangled connections; then the $\Pr \left({\rm {\mathcal P}}^{S} \right)$ probability that a given single path ${\rm {\mathcal P}}^{S} $ exists between $A_{\rho ,U_{k} } $ and $B_{\rho ,U_{k} } $with the fidelity criterion is expressed as
\begin{equation} \label{ZEqnNum467938} 
\begin{split}
   \Pr \left( {{\mathcal{P}}^{S}} \right)&=\prod\limits_{{{E}_{{{\text{L}}_{l}}}}\left( x,y \right)\in {{\mathcal{P}}^{S}}}{\Pr \left( {{F}_{\Delta }}\left( x,y \right)<{{{\hat{F}}}_{\Delta }} \right)} \\ 
 & ={{\left( \int\limits_{0}^{{{{\hat{F}}}_{\Delta }}}{\delta \left( z \right)dz} \right)}^{g}}.  
\end{split}
\end{equation}

\subsubsection{Multipath Entanglement Accessibility}

Let ${\rm {\mathcal P}}_{i}^{M} $, $i=1,\ldots ,m$, refer to the $i$th multipath between $A_{\rho ,U_{k} } $ and $B_{\rho ,U_{k} } $, which means that entanglement can be distributed from $A_{\rho ,U_{k} } $ to $B_{\rho ,U_{k} } $ through a set ${\rm {\mathcal P}}_{M} $ of $m$ end-to-end connection-disjoint entangled paths as ${\rm {\mathcal P}}_{M} =\left\{{\rm {\mathcal P}}_{1}^{M} ,\ldots ,{\rm {\mathcal P}}_{m}^{M} \right\}$ in the network $N$. The $\Pr \left({\rm {\mathcal P}}_{M} \right)$ probability \cite{ref61} that  $A_{\rho ,U_{k} } $ and $B_{\rho ,U_{k} } $ share a common entanglement with the fidelity criterion is as
\begin{equation} \label{ZEqnNum653831} 
\Pr \left({\rm {\mathcal P}}_{M} \right)=1-\prod _{S=1}^{m}\left(1-\Pr \left({\rm {\mathcal P}}^{S} \right)\right)=1- \prod _{S=1}^{m}\left(1-\left(\int\limits_{0}^{\hat{F}_{\Delta } }\delta \left(z\right)dz \right)^{g_{S} } \right) ,                  
\end{equation} 
where $S$ is a path index, and $g_{S} $ is the number of entangled connections associated with ${\rm {\mathcal P}}^{S} $.

Based on the distribution of $F_{\Delta } \left(x,y\right)$ fidelity distances between the node pairs of the network, the formulas of \eqref{ZEqnNum467938} and \eqref{ZEqnNum653831} can be derived in a more exact form\footnote{Assuming an exponential distribution of $F_{\Delta } \left(x,y\right)$, $\Pr \left({\rm {\mathcal P}}^{S} \right)=\int\limits_{0}^{\hat{F}_{\Delta } }\lambda e^{-\lambda z}  dz=1-e^{-\lambda \hat{F}_{\Delta } }$, where $\lambda $ is a distribution coefficient, while $\Pr \left({\rm {\mathcal P}}_{M} \right)=1-\prod _{S=1}^{m}\left(1-\left(1-e^{-\lambda \hat{F}_{\Delta } } \right)^{g_{S} } \right)$.}.

\section{Integrated Criterion on Connection Probability and Fidelity}
\label{sec4}
The integrated criterion extends the results of \sref{sec3} to include the criterion on the probability of the existence of a given entangled connection between a node pair of a path. Using the integrated criterion on the connection probability and entanglement fidelity, we derive the probability of existence of single-path and multipath sets with $m$ end-to-end connection-disjoint entangled paths \cite{ref61} between source and target nodes.

In our model, the fidelity of shared entanglement evolves in time for a given node pair $\left(x,y\right)$. For each quantum node $i$ at a time $t$, let $\Psi _{i} \left(t\right)$ be defined as 
\begin{equation} \label{7)} 
\Psi _{i} \left(t\right)=\left(\Pr \left(E_{{\rm L}_{l} } \left(i,j,t\right)\right),F_{i} \left(t\right)\right)^{T} , 
\end{equation} 
where $\Pr \left(E_{{\rm L}_{l} } \left(i,j,t\right)\right)$ is the probability of an ${\rm L}_{l} $-level entangled connection with a node $j$ determined in node $i$ at a time $t$, while $F_{i} \left(t\right)$ is the fidelity of entanglement determined in node $i$ at a time $t$. 

For a node pair $\left(x,y\right)$, according to local quantities, the following distance can be defined:
\begin{equation} \label{ZEqnNum625413} 
\Delta \left(\Pr \left(E_{{\rm L}_{l} } \left(t\right)\right)\right)=\left|\Pr \left(E_{{\rm L}_{l} } \left(x,y,t\right)\right)-\Pr \left(E_{{\rm L}_{l} } \left(y,x,t\right)\right)\right|,                          
\end{equation} 
where $\Pr \left(E_{{\rm L}_{l} } \left(x,y,t\right)\right)$, and $\Pr \left(E_{{\rm L}_{l} } \left(y,x,t\right)\right)$ are the connection probability quantities determined in nodes $x$ and $y$, and the fidelity distance $F_{\Delta } \left(t\right)$ is described by
\begin{equation} \label{ZEqnNum771887} 
F_{\Delta } \left(t\right)=\left|F_{x} \left(t\right)-F_{y} \left(t\right)\right|,                                             
\end{equation} 
where $F_{x} \left(t\right)$, and $F_{y} \left(t\right)$ are the fidelity quantities determined in nodes $x$ and $y$. 

A distance of $\Psi _{x} \left(t\right)$ and $\Psi _{y} \left(t\right)$ for a node pair $\left(x,y\right)$ at a particular time $t$ is expressed via $\gamma _{x,y} \left(t\right)$, as 
\begin{equation} \label{ZEqnNum810096} 
\begin{split}
   {{\gamma }_{x,y}}\left( t \right)&=\left| {{\Psi }_{x}}\left( t \right)-{{\Psi }_{y}}\left( t \right) \right| \\ 
 & ={{\left( {{\left( \Pr \left( {{E}_{{{\text{L}}_{l}}}}\left( x,y,t \right) \right)-\Pr \left( {{E}_{{{\text{L}}_{l}}}}\left( y,x,t \right) \right) \right)}^{2}}+{{\left( {{F}_{x}}\left( t \right)-{{F}_{y}}\left( t \right) \right)}^{2}} \right)}^{\frac{1}{2}}}.  
\end{split}
\end{equation} 
Since the connection probability and the entanglement fidelity parameters evolve over time, after $\Delta t$ from an initial time $t_{0} $, the quantity $\Psi _{x} \left(t_{0} +\Delta t\right)$ of a given node $x$ evolves as
\begin{equation} \label{ZEqnNum556643} 
{{\Psi }_{x}}\left( {{t}_{0}}+\Delta t \right)={{\Psi }_{x}}\left( {{t}_{0}} \right)+{{\chi }_{x}}\left( {{t}_{0}},\Delta t \right),
\end{equation} 
where $\chi _{x} \left(t_{0} ,\Delta t\right)$ is expressed as
\begin{equation} \label{ZEqnNum698129} 
{{\chi }_{x}}\left( {{t}_{0}},\Delta t \right)=\left( \begin{matrix}
\begin{split}
 & \chi _{x}^{\Pr \left( {{E}_{{{\text{L}}_{l}}}}\left( x,y \right) \right)}\left( {{t}_{0}},\Delta t \right)  \\
& \chi _{x}^{{{F}_{x}}}\left( {{t}_{0}},\Delta t \right)\text{         }  \\
\end{split}
\end{matrix} \right)=\left( \begin{matrix}
\begin{split}
  & \int\limits_{{{t}_{0}}}^{{{t}_{0}}+\Delta t}{\phi _{x}^{\Pr \left( {{E}_{{{\text{L}}_{l}}}}\left( x,y \right) \right)}\left( q \right)dq}  \\
  & \int\limits_{{{t}_{0}}}^{{{t}_{0}}+\Delta t}{\phi _{x}^{{{F}_{x}}}\left( q \right)dq\text{          }}  \\
\end{split}
\end{matrix} \right),
\end{equation} 
where $\delta \left(\gamma _{x,y} \right)$ is the probability density function of distance function $\gamma _{x,y} $, and $\phi _{x}^{\Pr \left(E_{{\rm L}_{l} } \left(x,y\right)\right)} $ and $\phi _{x}^{F_{x} } $ are expressed as the connection probability and entanglement fidelity evolution functions of node $x$. 

For a given node pair $\left(x,y\right)$, the particular upper bound $\gamma _{x,y}^{\max } $ on the maximal allowable distance between $\Psi _{x} \left(t_{0} +\Delta t\right)$ and $\Psi _{y} \left(t_{0} +\Delta t\right)$ at time $t_{0} +\Delta t$ leads to a limit while $\left(x,y\right)$ can be referred to as entangled:
\begin{equation} \label{ZEqnNum753073} 
\begin{split}
   {{\gamma }_{x,y}}\left( {{t}_{0}}+\Delta t \right)&=\left| {{\Psi }_{x}}\left( {{t}_{0}}+\Delta t \right)-{{\Psi }_{y}}\left( {{t}_{0}}+\Delta t \right) \right| \\ 
 & =\left| {{\Psi }_{x}}\left( {{t}_{0}} \right)+{{\chi }_{x}}\left( {{t}_{0}},\Delta t \right)-{{\Psi }_{y}}\left( {{t}_{0}} \right)-{{\chi }_{y}}\left( {{t}_{0}},\Delta t \right) \right|\le \gamma _{x,y}^{\max }.  
\end{split}
\end{equation} 
If $\gamma _{x,y} \left(t_{0} +\Delta t\right)$ exceeds $\gamma _{x,y}^{\max } $, then the difference of the local entangled connection probabilities and entanglement fidelities are above a critical limit; therefore, the node pair $\left(x,y\right)$ is referred to as unentangled. 

\begin{center}
\begin{figure*}[h!]
%\vspace{-0.4cm}
\begin{center}
\includegraphics[angle = 0,width=1\linewidth]{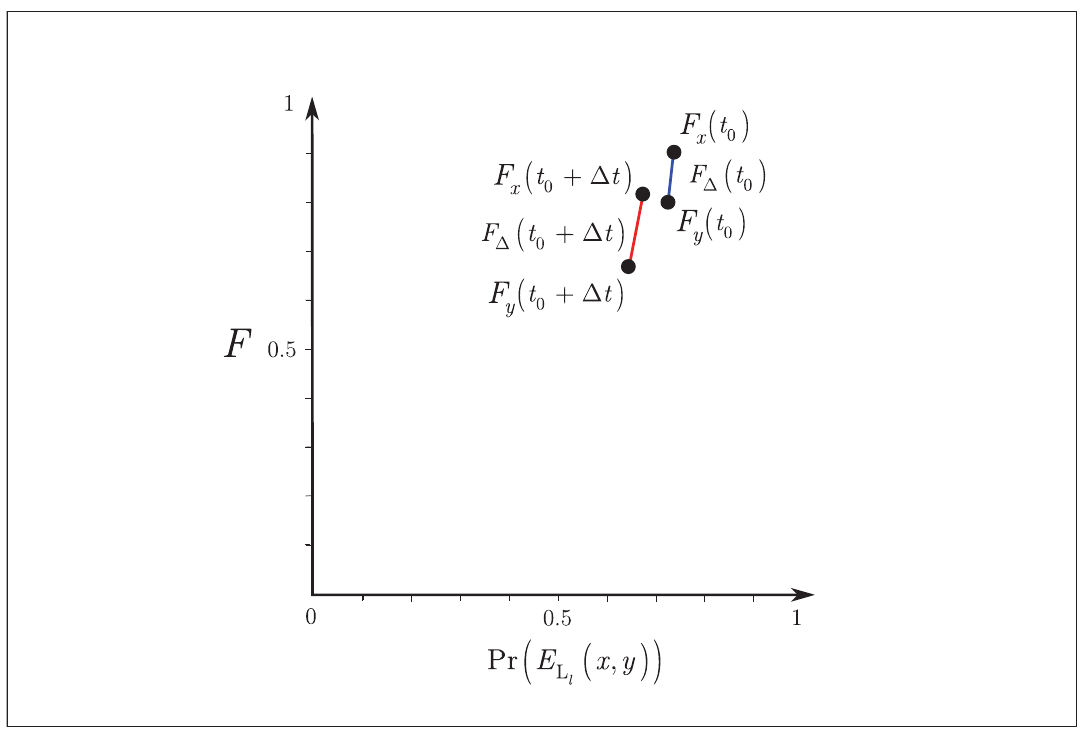}
\caption{Evolution of $F_{\Delta } \left(t\right)$ and $\Pr \left(E_{{\rm L}_{l} } \left(x,y\right)\right)$ for a node pair $\left(x,y\right)$ at $t=t_{0} $ and $t=t_{0} +\Delta t$. $F_{\Delta } \left(t_{0} \right)$ is $\left|F_{x} \left(t_{0} \right)-F_{y} \left(t_{0} \right)\right|$, where $F_{x} \left(t_{0} \right),F_{y} \left(t_{0} \right)$ are the fidelities of shared entanglement in the nodes, $F_{\Delta } \left(t_{0} +\Delta t\right)$ yields the difference $\left|F_{x} \left(t_{0} +\Delta t\right)-F_{y} \left(t_{0} +\Delta t\right)\right|$.} 
 \label{fig1}
 \end{center}
\end{figure*}
\end{center}

Using \eqref{ZEqnNum810096} and \eqref{ZEqnNum698129}, \eqref{ZEqnNum753073} can be rewritten as
\begin{equation} \label{14)} 
\begin{split}
   {{\gamma }_{x,y}}\left( {{t}_{0}}+\Delta t \right)=&\left( {{\left( \Pr \left( {{E}_{{{\text{L}}_{l}}}}\left( x,y,{{t}_{0}}+\Delta t \right) \right)-\Pr \left( {{E}_{{{\text{L}}_{l}}}}\left( y,x,{{t}_{0}}+\Delta t \right) \right) \right)}^{2}} \right. \\ 
 & {{\left. +{{\left( {{F}_{x}}\left( {{t}_{0}}+\Delta t \right)-{{F}_{y}}\left( {{t}_{0}}+\Delta t \right) \right)}^{2}} \right)}^{\frac{1}{2}}},  
\end{split}
\end{equation} 
which leads to
\begin{equation} \label{ZEqnNum750524} 
\begin{split}
  & {{\gamma }_{x,y}}\left( {{t}_{0}}+\Delta t \right) \\ 
  =&\left( {{\left( \Pr \left( {{E}_{{{\text{L}}_{l}}}}\left( x,y,{{t}_{0}} \right) \right)+\int\limits_{{{t}_{0}}}^{{{t}_{0}}+\Delta t}{\phi _{x}^{\Pr \left( {{E}_{{{\text{L}}_{l}}}}\left( x,y \right) \right)}\left( q \right)dq-\left( \begin{split}
  & \Pr \left( {{E}_{{{\text{L}}_{l}}}}\left( y,x,{{t}_{0}} \right) \right) +\int\limits_{{{t}_{0}}}^{{{t}_{0}}+\Delta t}{\phi _{y}^{\Pr \left( {{E}_{{{\text{L}}_{l}}}}\left( y,x \right) \right)}\left( q \right)dq} \\ 
\end{split} \right)} \right)}^{2}} \right. \\ 
 & {{\left. +{{\left( {{F}_{x}}\left( {{t}_{0}} \right)+\int\limits_{{{t}_{0}}}^{{{t}_{0}}+\Delta t}{\phi _{x}^{{{F}_{x}}}\left( q \right)dq-\left( \begin{split}
  & {{F}_{y}}\left( {{t}_{0}} \right) +\int\limits_{{{t}_{0}}}^{{{t}_{0}}+\Delta t}{\phi _{y}^{{{F}_{y}}}\left( q \right)dq} \\ 
\end{split} \right)} \right)}^{2}} \right)}^{\frac{1}{2}}}.  
\end{split}
\end{equation} 
A representation of $F_{\Delta } \left(t_{0} \right)$ and $F_{\Delta } \left(t_{0} +\Delta t\right)$ for a node pair $\left(x,y\right)$ is depicted in \fref{fig1}. The $\Pr \left(E_{{\rm L}_{l} } \left(x,y\right)\right)$ connection probability is assumed to be different in the nodes at a particular time. 

\subsection{Single-Path Entanglement Accessibility}

After these derivations, the $\Pr \left({\rm {\mathcal P}}^{S} \left(t_{0} +\Delta t\right)\right)$ probability of a ${\rm {\mathcal P}}^{S} $ single path in function of the connection probability and entanglement fidelity in the nodes of the path (e.g., connection probability criterion and fidelity criterion for all entangled connections of the path) is as follows.

Using $\gamma _{x,y} \left(t_{0} +\Delta t\right)$ in \eqref{ZEqnNum750524}, the probability of the existence of a given single path ${\rm {\mathcal P}}^{S} $ with ${{g}_{S}}$ entangled connections between $A_{\rho ,U_{k} } $ and $B_{\rho ,U_{k} } $ with a connection probability criterion and fidelity criterion for all entangled connections (see \eqref{ZEqnNum467938}) at time $t_{0} +\Delta t$ can therefore be rewritten as 
\begin{equation} \label{16)} 
\begin{split}
  & \Pr \left( {{\mathcal{P}}^{S}}\left( {{t}_{0}}+\Delta t \right) \right) \\ 
 & =\prod\limits_{{{E}_{{{\text{L}}_{l}}}}\left( x,y \right)\in {{\mathcal{P}}^{S}}}{\Pr \left( {{\gamma }_{x,y}}\left( {{t}_{0}}+\Delta t \right)<\gamma _{x,y}^{\max } \right)} \\ 
 & ={{\left( \int\limits_{0}^{\gamma _{x,y}^{\max }}{\delta \left( q \right)dq} \right)}^{{{g}_{S}}}},  
\end{split}
\end{equation} 
where
\begin{equation} \label{17)} 
\Pr \left(\gamma _{x,y} \left(t_{0} +\Delta t\right)<\gamma _{x,y}^{\max } \right)=\int\limits_{0}^{\gamma _{x,y}^{\max } }\delta \left(q\right)dq .                                
\end{equation}

\subsection{Multipath Entanglement Accessibility}

For the multipath scenario, \eqref{ZEqnNum653831} can be written via \eqref{ZEqnNum750524} as follows. For a given set of $m$ end-to-end connection-disjoint entangled paths expressed as ${\rm {\mathcal P}}_{M} =\left\{{\rm {\mathcal P}}_{1}^{M} ,\ldots ,{\rm {\mathcal P}}_{m}^{M} \right\}$ between $A_{\rho ,U_{k} } $ and $B_{\rho ,U_{k} } $, the $\Pr \left({\rm {\mathcal P}}_{M} \left(t_{0} +\Delta t\right)\right)$ probability that  $A_{\rho ,U_{k} } $ and $B_{\rho ,U_{k} } $ share a common entanglement with a connection probability criterion and fidelity criterion at time $t_{0} +\Delta t$ is expressed as
\begin{equation} \label{18)} 
\begin{split}
   \Pr \left( {{\mathcal{P}}_{M}}\left( {{t}_{0}}+\Delta t \right) \right)&=1-\prod\limits_{S=1}^{m}{\left( 1-\Pr \left( {{\mathcal{P}}^{S}}\left( {{t}_{0}}+\Delta t \right) \right) \right)} \\ 
 & =1-\prod\limits_{S=1}^{m}{\left( 1-{{\left( \int\limits_{0}^{\gamma _{x,y}^{\max }}{\delta \left( q \right)dq} \right)}^{{{g}_{S}}}} \right)}.  
\end{split}
\end{equation} 

\section{Control of Entanglement Access}
\label{sec5}
The entanglement access control algorithm establishes a number of connection-disjoint entangled paths between a source node and a target node. Changing the number $m$ of connection-disjoint entangled paths allows us to modify both the probability of entanglement between the source and target nodes and the fidelity of available entanglement in the end nodes. 

For the algorithm, a $c\left(E_{{\rm L}_{l} } \left(x,y\right)\right)$ cost function \cite{ref61} of a given entangled connection $E_{{\rm L}_{l} } \left(x,y\right)$ is defined as 
\begin{equation} \label{ZEqnNum323728} 
\begin{split}
   c\left( {{E}_{{{\text{L}}_{l}}}}\left( x,y \right) \right)=&\left( {{\left( \chi _{x}^{\Pr \left( {{E}_{{{\text{L}}_{l}}}}\left( x,y \right) \right)}\left( {{t}_{0}},\Delta t \right)-\chi _{y}^{\Pr \left( {{E}_{{{\text{L}}_{l}}}}\left( y,x \right) \right)}\left( {{t}_{0}},\Delta t \right) \right)}^{2}} \right. \\ 
 & {{\left. +{{\left( \chi _{x}^{{{F}_{x}}}\left( {{t}_{0}},\Delta t \right)-\chi _{y}^{{{F}_{y}}}\left( {{t}_{0}},\Delta t \right) \right)}^{2}} \right)}^{\frac{1}{2}}},  
\end{split}
\end{equation} 
where 
\begin{equation} \label{20)}
 \chi _{x}^{\Pr \left( {{E}_{{{\text{L}}_{l}}}}\left( x,y \right) \right)}\left( {{t}_{0}},\Delta t \right)=\int\limits_{{{t}_{0}}}^{{{t}_{0}}+\Delta t}{\phi _{x}^{\Pr \left( {{E}_{{{\text{L}}_{l}}}}\left( x,y \right) \right)}\left( q \right)dq},
\end{equation}
\begin{equation}
\chi _{x}^{{{F}_{x}}}\left( {{t}_{0}},\Delta t \right)\text{=}\int\limits_{{{t}_{0}}}^{{{t}_{0}}+\Delta t}{\phi _{x}^{{{F}_{x}}}\left( q \right)dq,}\text{      }  
\end{equation}
and
\begin{equation} \label{21)}
\chi _{y}^{\Pr \left( {{E}_{{{\text{L}}_{l}}}}\left( y,x \right) \right)}\left( {{t}_{0}},\Delta t \right)=\int\limits_{{{t}_{0}}}^{{{t}_{0}}+\Delta t}{\phi _{y}^{\Pr \left( {{E}_{{{\text{L}}_{l}}}}\left( y,x \right) \right)}\left( q \right)dq},
\end{equation}
\begin{equation}\label{endeq}
\chi _{y}^{{{F}_{y}}}\left( {{t}_{0}},\Delta t \right)\text{=}\int\limits_{{{t}_{0}}}^{{{t}_{0}}+\Delta t}{\phi _{y}^{{{F}_{y}}}\left( q \right)dq.}  
\end{equation}

Let $N$ be the actual quantum repeater network with $\left|V\right|$ quantum nodes. A given ${\rm L}_{l} $-level entangled connection between a node pair $\left(x,y\right)$ is expressed as $E_{{\rm L}_{l} } \left(x,y\right)$. 

Let $A_{\rho ,U_{k}} $ and $B_{\rho ,U_{k}} $ be the source and target quantum nodes of a demand $\rho $ associated with user $U_{k} $. Using \eqref{ZEqnNum323728} and a given entangled path ${\rm {\mathcal P}}$ with a set of $q$ quantum repeaters $R_i$, $i=1,\ldots ,q$, and a set $\mathcal{S}$ entangled connections, as
\begin{equation} \label{22)} 
\mathcal{S}=\left\{ {{E}_{{{\text{L}}_{l}}}}\left( {{A}_{\rho ,U_{k}}},R_1 \right),\ldots ,{{E}_{{{\text{L}}_{l}}}}\left(R_q,{{B}_{\rho ,U_{k}}} \right) \right\},
\end{equation} 
the cost of path ${\rm {\mathcal P}}$ is defined as 
\begin{equation} \label{ZEqnNum732036} 
c\left({\rm {\mathcal P}}\right)=c\left(E_{{\rm L}_{l} } \left(A_{\rho ,U_{k}}, R_1\right)\right)+\ldots +c\left(E_{{\rm L}_{l} } \left(R_q,B_{\rho ,U_{k}} \right)\right).                              
\end{equation} 
The ${\rm {\mathcal D}}_{{\rm {\mathcal A}}} $ entanglement access control algorithm outputs a set of ${\rm {\mathcal P}}_{M} =\left\{{\rm {\mathcal P}}_{1}^{M} ,\ldots ,{\rm {\mathcal P}}_{m}^{M} \right\}$, which contains the $m$ connection-disjoint entangled paths between  $A_{\rho ,U_{k}} $ and $B_{\rho ,U_{k}} $. 

In function of $m$, ${\rm {\mathcal U}}_{C} $ priority classes can be defined for the users of the quantum Internet. A high priority user gets a high value of $m$, while lower priority users get lower values of $m$. The actual value of $m$ for a particular user class ${\rm {\mathcal U}}_{C} $ can be determined in function of the current network conditions.

The steps are given in Algorithm 1.

 \setcounter{algocf}{0}
\begin{algo}
  \DontPrintSemicolon
\caption{Entanglement access control in the quantum Internet}

\textbf{Step 1} At a given ${\rm {\mathcal U}}_{C}({U_k}) $ priority class of user $U_k$, set the number $m$ of accessible entangled paths for a particular demand $\rho $ of a given user $U_k$. 

\textbf{Step 2}. For the given demand $\rho $, establish $m$ connection-disjoint entangled connections from $A_{\rho, U_k} $ with the direct neighbor nodes of $A_{\rho, U_k} $ in the following manner.

\textbf{Step 3}. For all entangled connections of $A_{{\rho, U_k} } $, determine the path cost $c\left({\rm {\mathcal P}}_{i}^{M} \right)$, $i=1,\ldots ,m$ via \eqref{ZEqnNum732036} using the entangled connection cost $c\left(E_{{\rm L}_{l} } \left(A_{{\rho, U_k} } ,{R_{i}} \right)\right)$ from \eqref{ZEqnNum323728}, where ${R_{i}} $ is a quantum repeater node.

\textbf{Step 4}. For all next neighbor nodes $R_j$ of quantum repeater ${R_{i}} $, establish entanglement from quantum repeater ${R_{i}} $ to quantum repeater $R_j$. Compute $\chi _{{R_{i}} }^{\Pr \left(E_{{\rm L}_{l} } \left({R_{i}} ,R_j\right)\right)} \left(t_{0} ,\Delta t\right),$ $\chi _{R_j}^{\Pr \left(E_{{\rm L}_{l} } \left(R_j,{R_{i}} \right)\right)} \left(t_{0} ,\Delta t\right),$ $\chi _{{R_{i}} }^{F_{{R_{i}} } } \left(t_{0} ,\Delta t\right)$ and $\chi _{R_j}^{F_{R_j} } \left(t_{0} ,\Delta t\right)$ via \eqref{20)}-\eqref{endeq}, and increase the $c\left({\rm {\mathcal P}}_{i}^{M} \right)$ path cost by $c\left(E_{{\rm L}_{l} } \left({R_{i}} ,R_j\right)\right)$. 

\textbf{Step 5}. If quantum repeater $R_j$ has no entangled connections with ${R_{i}} $, then establish entanglement with a different neighbor $R_k $ of $R_j$ from ${R_{i}} $ towards $B_{{\rho, U_k} } $. Compute $\chi _{{R_{i}} }^{\Pr \left(E_{{\rm L}_{l} } \left({R_{i}} ,R_k \right)\right)} \left(t_{0} ,\Delta t\right),$ $\chi _{R_k }^{\Pr \left(E_{{\rm L}_{l} } \left(R_k ,{R_{i}} \right)\right)} \left(t_{0} ,\Delta t\right),$ $\chi _{{R_{i}} }^{F_{{R_{i}} } } \left(t_{0} ,\Delta t\right)$ and $\chi _{R_k }^{F_{R_k } } \left(t_{0} ,\Delta t\right)$ via \eqref{20)}-\eqref{endeq}, and increase the $c\left({\rm {\mathcal P}}_{i}^{M} \right)$ path cost by $c\left(E_{{\rm L}_{l} } \left({R_{i}} ,R_k \right)\right)$.

\textbf{Step 6}. Repeat the steps until $B_{{\rho, U_k} } $ is reached. Output set ${\rm {\mathcal P}}_{M} =\left\{{\rm {\mathcal P}}_{1}^{M} ,\ldots ,{\rm {\mathcal P}}_{m}^{M} \right\}$ and the path costs $c\left({\rm {\mathcal P}}_{i}^{M} \right)$ for all paths of ${\rm {\mathcal P}}_{M}$.

\textbf{Step 7}. Evaluate the $\Pr \left({\rm {\mathcal P}}_{M} \right)$ probability for user $U_k$ via \eqref{18)}. If $\Pr \left({\rm {\mathcal P}}_{M} \right) < \Pr _{{{U}_{k}}}^{*}\left( {{\mathcal{P}}_{M}} \right)$, where $\Pr _{{{U}_{k}}}^{*}\left( {{\mathcal{P}}_{M}} \right)$ is the critical lower bound on $\Pr \left({\rm {\mathcal P}}_{M} \right)$ set for ${U_k}$, then increase $m$, $m=m+1$. If $\Pr \left({\rm {\mathcal P}}_{M} \right) \ge \left\langle {{\Pr }_{{{U}_{k}}}}\left( {{\mathcal{P}}_{M}} \right) \right\rangle $, where $\left\langle {{\Pr }_{{{U}_{k}}}}\left( {{\mathcal{P}}_{M}} \right) \right\rangle $ is the maximal allowed value of $\Pr \left({\rm {\mathcal P}}_{M} \right)$ for ${U_k}$, then decrease $m$, $m=m-1$. If $\Pr _{{{U}_{k}}}^{*}\left( {{\mathcal{P}}_{M}} \right) \le \Pr \left({\rm {\mathcal P}}_{M} \right) < \left\langle {{\Pr }_{{{U}_{k}}}}\left( {{\mathcal{P}}_{M}} \right) \right\rangle $, then leave $m$ unchanged.

\textbf{Step 8}. Repeat steps 1-7 for all ${U_k}$, $k=1,,\ldots K$.

\end{algo} 

\subsection{Description}

A brief description of the ${\rm {\mathcal D}}_{{\rm {\mathcal A}}} $ entanglement access control algorithm is as follows. 

Step 1 sets $m$ for a user $U_k$ by the ${\rm {\mathcal U}}_{C}({U_k}) $ priority class of the user. The ${\rm {\mathcal U}}_{C}({U_k}) $ determines the available value(s) of $m$ for $U_k$.

In Step 2, entanglement is established between the source node $A_{\rho ,U_{k} } $ of the given demand of the user and the neighboring quantum repeaters. The relevant metrics quantities are also calculated in this step. 

Using the derived quantities of Step 2, in Step 3, the cost paths are derived via \eqref{ZEqnNum732036} using the entangled connection cost formula of \eqref{ZEqnNum323728}. 

Steps 4-5 deal with the intermediate quantum repeater nodes associated with the given demand. These steps also ensure that entanglement is distributed through the cheapest path $c\left({\rm {\mathcal P}}'\right)$ from a source node $A_{\rho ,U_{k} } $ towards $B_{\rho ,U_{k} } $, via a given intermediate repeater node $R_i $. It is ensured in our model that if the intermediate repeater node $R_i$ also shares entanglement with a quantum repeater $R_j$, then node $R_j$ will not establish entanglement with $B_{\rho ,U_{k} } $, since $B_{\rho ,U_{k} } $ can be reached via $R_i$, which is on the cheapest path $c\left({\rm {\mathcal P}}'\right)$.

Step 6 outputs the set of $m$ end-to-end connection-disjoint entangled paths between $A_{\rho ,U_{k} } $ and $B_{\rho ,U_{k} } $ and the path costs for all paths.

Step 7 determines the $\Pr \left({\rm {\mathcal P}}_{M} \right)$ probability for user $U_k$ via \eqref{18)}, and updates the actual value of $m$ if needed.

Finally, Step 8 extends the steps for all users. 

\begin{center}
\begin{figure*}[h!]
%\vspace{-0.4cm}
\begin{center}
\includegraphics[angle = 0,width=1\linewidth]{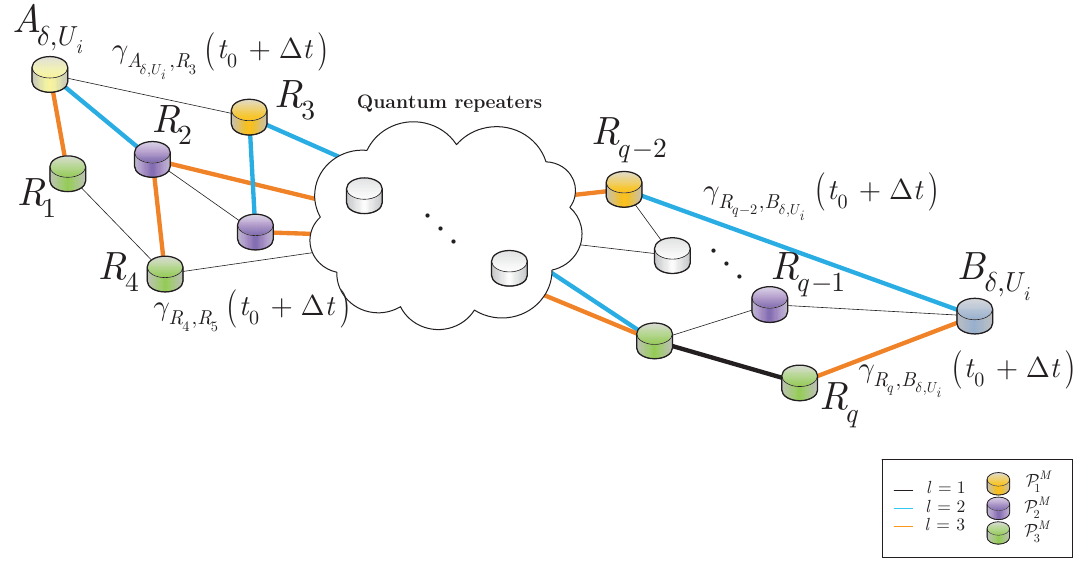}
\caption{A quantum Internet setting with $m=3$ connection-disjoint entangled paths ${\rm {\mathcal P}}_{1}^{M} $, ${\rm {\mathcal P}}_{2}^{M} $ and ${\rm {\mathcal P}}_{3}^{M} $ between an $i$th source quantum node $A_{\delta ,U_{i} } $ and target quantum node $B_{\delta ,U_{i} } $ with demand $\rho $, and $q$ intermediate $R_{i} $ quantum repeaters, $i=1,\ldots ,q$. The entangled paths consist of  $l=1$ level (direct) and multilevel, $l=2,3$ level entangled connections. The quantum nodes of ${\rm {\mathcal P}}_{1}^{M} $ are depicted by orange, the nodes of ${\rm {\mathcal P}}_{2}^{M} $ by purple, and the nodes of ${\rm {\mathcal P}}_{3}^{M} $ by green. The $\gamma _{x,y} \left(t_{0} +\Delta t\right)$ coefficients are derived for all entangled connections of the paths.} 
 \label{fig2}
 \end{center}
\end{figure*}
\end{center}

In \fref{fig2} a multipath entanglement accessibility is depicted in a quantum Internet setting with heterogeneous entanglement levels. The network situation depicts connection-disjoint entangled paths that share no common entangled connection between a source node $A_{\delta ,U_{i} } $ and a receiver node $B_{\delta ,U_{i} } $. The entangled paths are characterized by the derived formulas.

\subsection{Computational Complexity}
For a given quantum network $N$ with $\left|V\right|$ quantum nodes, the computational complexity of the ${\rm {\mathcal D}}_{{\rm {\mathcal A}}} $ entanglement access control algorithm for a given demand $\rho $ is at most ${\rm {\mathcal O}}\left(\left|V\right|\right)$, since the problem is analogous to the establishment of a path by message broadcasting \cite{ref61}.

\subsection{Numerical Evidence}
We provide a numerical evidence on the distribution of the ${\rm {\mathcal P}}^{S} $ and ${\rm {\mathcal P}}_{M} $ path probabilities. 

Let us set $F_{crit} =0.98$ for the lower bound on the fidelity of entanglement between all node pairs $x$ and $y$, $F_{x} \ge F_{crit} $, and $F_{y} \ge F_{crit} $. Then, the maximal allowed fidelity distance is set as $\hat{F}_{\Delta } =1-0.98=0.02$. 

Then, let us assume that a ${\rm {\mathcal P}}^{S} $ single path between $A_{\rho ,U_{k} } $ and $B_{\rho ,U_{k} } $ consist of $g$ entangled connections with different $l$ entanglement levels between the nodes of the path ${\rm {\mathcal P}}^{S} $. For the ${\rm {\mathcal P}}_{M} =\left\{{\rm {\mathcal P}}_{1}^{M} ,\ldots ,{\rm {\mathcal P}}_{m}^{M} \right\}$ multipath scenario, each entangled path consist of $g_{S} $ entangled connections with different $l$ entanglement levels between the nodes of each entangled path of ${\rm {\mathcal P}}_{M} $.

\begin{center}
\begin{figure*}[h!]
%\vspace{-0.4cm}
\begin{center}
\includegraphics[angle = 0,width=1\linewidth]{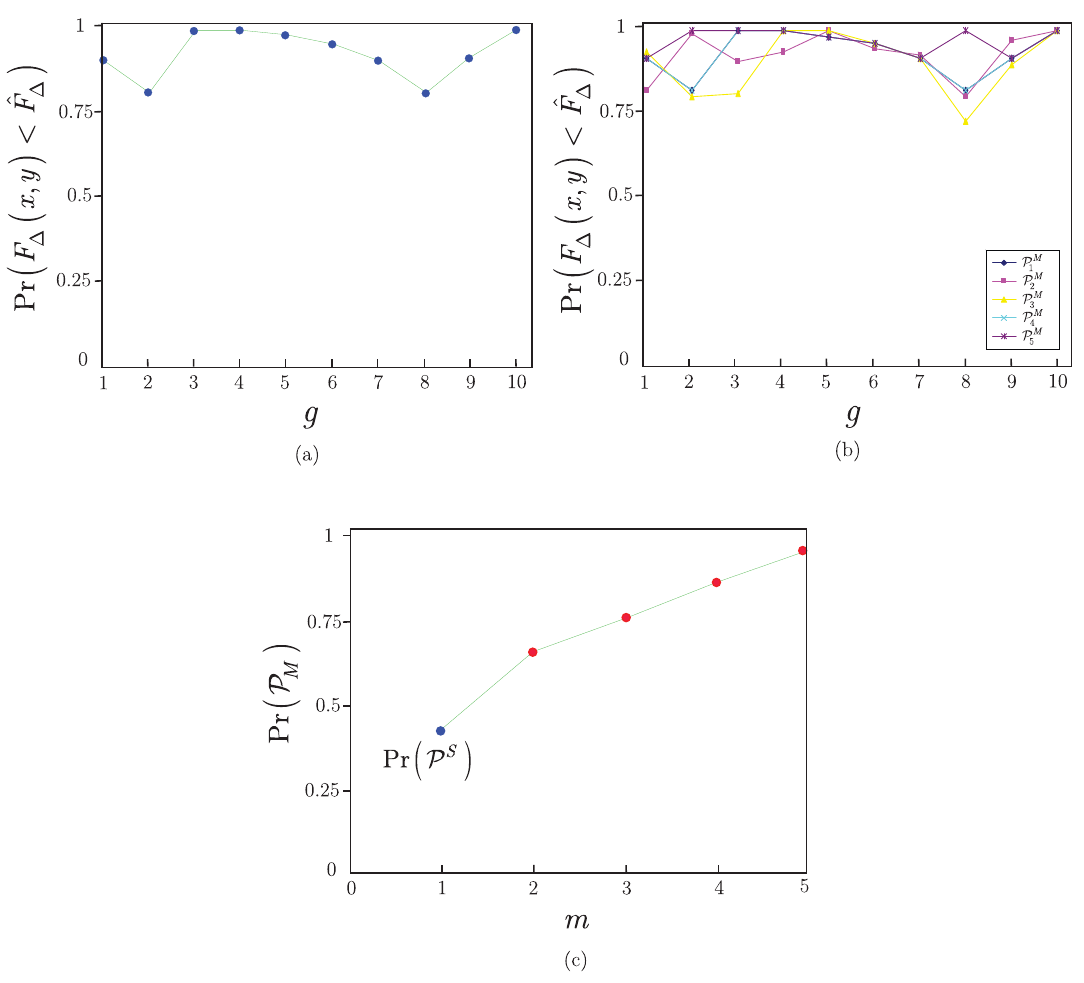}
\caption{The probability of shared entanglement between the source and the destination with the fidelity criterion at a single entangled path and at a multipath setting, $F_{crit} =0.98$, $\hat{F}_{\Delta } =0.02$, $g_{S} =g=10$. (a): A distribution of the $\Pr \left(F_{\Delta } \left(x,y\right)<\hat{F}_{\Delta } \right)$ probabilities for the $g$  entangled connections of a ${\rm {\mathcal P}}^{S} $ single path. (b): A distribution of the $\Pr \left(F_{\Delta } \left(x,y\right)<\hat{F}_{\Delta } \right)$ probabilities for the ${g_{S}}=g$  entangled connections of a ${\rm {\mathcal P}}_{M} =\left\{{\rm {\mathcal P}}_{1}^{M} ,\ldots ,{\rm {\mathcal P}}_{m}^{M} \right\}$ multipath setting, $m=5$. (c): The $\Pr \left({\rm {\mathcal P}}_{M} \right)$ probabilities in function of $m$. The $\Pr \left({\rm {\mathcal P}}_{S} \right)$ single path probability is yielded at  $m=1$.} 
 \label{fig3}
 \end{center}
\end{figure*}
\end{center}

For simplicity let us assume that the number of entangled connections is set as $g_{S} =g=10$ for all entangled paths, and the distribution of the $\Pr \left(F_{\Delta } \left(x,y\right)<\hat{F}_{\Delta } \right)$ probabilities for the entangled connections of a ${\rm {\mathcal P}}^{S} $ single path at $\hat{F}_{\Delta } =0.02$ is as depicted in \fref{fig3}(a). The distribution of the $\Pr \left(F_{\Delta } \left(x,y\right)<\hat{F}_{\Delta } \right)$ probabilities for the entangled connections of a ${\rm {\mathcal P}}_{M} $ multipath with $m=5$ at $\hat{F}_{\Delta } =0.02$ is distributed as depicted in \fref{fig3}(b). The resulting $\Pr \left({\rm {\mathcal P}}^{S} \right)$ and $\Pr \left({\rm {\mathcal P}}_{M} \right)$ probabilities are depicted in \fref{fig3}(c).

The numerical analysis revealed that $\Pr \left({\rm {\mathcal P}}_{S} \right)\approx 0.4171$ for a ${\rm {\mathcal P}}_{S} $ single entangled path at the particular  $\Pr \left(F_{\Delta } \left(x,y\right)<\hat{F}_{\Delta } \right)$ connection-level values of the path (given in \fref{fig3}(a)). The ${\rm {\mathcal P}}_{M} $ multipath setting at connection-level values of \fref{fig3}(b), at $m=4$ doubles the success probability of the single path setting with $\Pr \left({\rm {\mathcal P}}_{M} \right)\approx 0.8549$, while at $m=5$, the resulting probability is $\Pr \left({\rm {\mathcal P}}_{M} \right)\approx 0.9476$.

\section{Conclusions}
\label{sec6}
In this work, we defined a method to achieve entanglement access control in the quantum Internet. The algorithm utilizes different paths between the source and target nodes in function of a particular path cost function. The path cost function uses the local entanglement fidelities of the nodes and the probability of the existence of the entangled connections. Increasing the number of available paths leads to a multipath setting, which allows the parties to establish high fidelity entanglement with reliable entangled connections between the end nodes. The proposed scheme has moderate complexity, and it is particularly convenient for the entangled quantum network structure of the quantum Internet.

\section*{Acknowledgements}
This work was partially supported by the National Research Development and Innovation Office of Hungary (Project No. 2017-1.2.1-NKP-2017-00001), by the Hungarian Scientific Research Fund - OTKA K-112125 and in part by the BME Artificial Intelligence FIKP grant of EMMI (BME FIKP-MI/SC).

%\section*{Statements}
%\subsection*{Ethics statement}
%This work did not involve any active collection of human data.
%\subsection*{Data accessibility statement}
%This work does not have any experimental data.
%\subsection*{Competing financial interests statement}
%We have no competing financial interests.
%\subsection*{Competing interests statement}
%We have no competing interests.
%\subsection*{Funding}
%No relevant funding. 
%\subsection*{Authors’ contributions}
%L.GY. designed the protocol and wrote the manuscript. L.GY. and S.I. analyzed the results. All authors reviewed the manuscript.

\end{document}